\documentclass[prb,twocolumn,floats,floatfix]{revtex4}

\usepackage{amsmath2000}
\usepackage{amsfonts}
\usepackage{epsfig}
\usepackage{bbm}

\newcommand{\eins}{\mathbbm{1}}

\begin{document}

\title{Decay Rate Distributions of Disordered Slabs and Application to
Random Lasers}

\author{M. Patra}
\affiliation{Instituut-Lorentz, Universiteit Leiden,
Postbus 9506, 2300 RA Leiden, The Netherlands}
\affiliation{Laboratory for Computational Engineering, 
Helsinki University of Technology, P.\,O. Box 9400, 02015 HUT, Finland}

\begin{abstract}
We compute the distribution of the decay rates (also referred to
as residues)
of the eigenstates of a disordered slab from a numerical
model. From the results of the numerical simulations, we are able to find simple 
analytical formulae that describe those results well. This is possible for
samples both in the diffusive and in the localised regime.
As example of a possible application, we investigate the lasing threshold of
random lasers.
\end{abstract}

\maketitle

\section{Introduction}

A very successful approach to describe disordered materials is supplied by
random-matrix theory, see Refs.~\onlinecite{beenakker:97a,guhr:98a} for reviews.
While one can put the beginning of random-matrix theory at Wigner's surmise for
describing the scattering spectra of heavy atomic nuclei,\cite{wigner:56a} its
theoretical foundations were laid only much later.\cite{bohigas:84a} It was very
successfully applied to  electronic transport in disordered wires and mesoscopic
quantum dots, and recently these methods have been adopted to model (quantum)
transport of optical radiation in media with spatially fluctuating dielectric
constant.\cite{beenakker:98a,beenakker:98b,patra:98a,patra:99a}

\begin{figure}
\epsfig{file=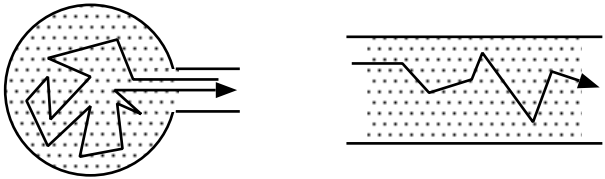,width=3.2in}
\caption{Two frequent geometries in the theory of disordered media are the
chaotic cavity (left) and the disordered slab (right). The motion in the chaotic
cavity is completely ergodic while in the disordered slab it is ergodic only in
the transversal direction.}
\label{figGeometrien}
\end{figure}

In the theoretical treatment of disordered materials, two particular geometries
are of special importance, namely the disordered slab and the chaotic cavity
(see Fig.~\ref{figGeometrien}). 
The principal difference between the geometries is easily explained: A chaotic
cavity is an object in which the dynamics is chaotic due to the shape of the
cavity or due to scatterers placed at random positions. The size of the opening
is small compared to the total surface area of the cavity. Particles
(electrons, photons) are then ``trapped'' inside the structure for a time that
is long enough to ergodically explore the entire cavity. In a disordered slab,
particles cannot be trapped that efficiently. They can no longer explore
the entire volume ergodically but they still stay long  enough to explore the
entire cross-section of the sample, thus still making a random-matrix
description possible. In order to call this geometry a ``wire''
or a ``waveguide'' its length should be much larger than its width. To be able
to apply the theory only the much weaker criterion that the length is sufficiently
larger
than the mean-free path of the medium has to be fulfilled.

Two different aspects are of special important in the theory of disordered
media, namely transport properties and resonances.
The transport properties (moments of the eigenvalues of the transmission and
reflection matrices) are known for the disordered slab in the limit that it is
wide,\cite{brouwer:98a} for the chaotic cavity with an opening that is so
small that only one or two modes can propagate through
it,\cite{brouwer:95b,brouwer:97a} or a chaotic cavity with a wide
opening.\cite{beenakker:98b}

Less is known about the poles of such systems. (The eigenvalues of the
Hamiltonian correspond to poles of the scattering matrix, and these show up as
resonances in a scattering experiment. Hence the somewhat
inconsistent nomenclature.) The beginning of random-matrix theory can be put at
the moment when Wigner surmised the eigenvalue distribution for a closed chaotic
cavity.\cite{wigner:56a,mehta:90,beenakker:97a} Here we are interested in open
systems, where the eigenvalues acquire an imaginary part. (The imaginary
part is referred to as
residue.) It determines the decay rate of the (quasi-)eigenstate of the
system. For chaotic cavities with 
broken time-reversal symmetry, the decay rate distribution is known
analytically for an opening of arbitrary size.\cite{fyodorov:97a} The
distribution for the more important case of preserved time-reversal
symmetry\footnote{Optical experiments always preserve time-reversal symmetry
unless a magneto-optical effect is included. For electric systems time-reversal
symmetry can be broken by applying a large magnetic field to the sample. (Such
fields are created routinely in experiments.)} is not known but can be
approximated by a cavity with broken symmetry and an opening of
half the real size.

Information on the residues of a disordered slab is very limited, and only the
scaling behaviour of the large residue-tail in the localised regime was
determined recently.\cite{titov:00a,terrano:00a} This deficiency is felt
especially strong in the random-laser community since the location of the
residues gives the lasing threshold of an optical system, and most experimental
setups resemble a disordered slab much more than they resemble a cavity. This
paper fills this gap by presenting the results of numerical simulations.  The
quality of the numerical decay rate distributions is good enough that it allows
us to arrive at analytical formulae for the distribution function, including
its dependence on the parameters of the system. The idea to use
high-quality simulations to arrive at formulae is not completely new as the
distribution of the scattering strengths of chaotic cavities was found in the
same way.\cite{beenakker:98b}

This paper is organised as follows: First, we introduce the Anderson Hamiltonian
used the describe the disordered slab. In Sec.~\ref{secEigenSolver} we show how
the eigenvalues of that Hamiltonian can be computed in a efficient numerical
way. Depending on the length of the slab, it can be in either the diffusive or
in the localised regime. We will analyse the decay rate distributions for both
regimes separately, first in Sec.~\ref{secDiffusive} for the diffusive and then
in Sec.~\ref{secLocalised} for the localised regime. Until that point all
results are completely general and can be applied to electronic and photonic
systems. In Sec.~\ref{secThreshold} we specialise on the lasing threshold in
amplifying disordered media. We distinguish between the diffusive and the
localised regimes (Secs.~\ref{secThresholdDiffusive}
and~\ref{secThresholdLocalised}). We conclude in Sec.~\ref{secConclusions}.

\section{Anderson Hamiltonian for a disordered slab}
\label{secModell}

We consider a two-dimensional slab of length $L$ and width $N$. The slab is
described by an Anderson-type lattice Hamiltonian with spacing $1$. 
In the Anderson model, transport is modelled by
nearest-neighbour hopping between lattice sites. Without
loss of generality we can set the hopping rate to $1$. With a
spatially varying potential $P(x,y)$ the Hamiltonian becomes
\begin{subequations}
\label{Hdef}
\begin{align}
\mathcal{H}_{(x,y),(x,y)} &= P(x,y) &(y \ne 1, L) \\
\mathcal{H}_{(x,y),(x,y)} &= P(x,y) - i \kappa &(y = 1, L) \label{Hdefkappa} \\
\mathcal{H}_{(x,y),(x+1,y)} &= 1 &(x<W) \\
\mathcal{H}_{(x,y),(x-1,y)} &= 1 &(x>1) \\
\mathcal{H}_{(x,y),(x,y+1)} &= 1 &(y< L) \\
\mathcal{H}_{(x,y),(x,y-1)} &= 1 &(y>1) 
\end{align}
\end{subequations}
All other elements are zero. $x$ runs from $1$ to $N$, and $y$ from $1$ to $L$. 

The real part $E$ of the eigenvalues of $\mathcal{H}$ in the limit of large $N$
and $L$ is confined to the interval $[-4;4]$.  (If the average of $P(x,y)$ is
nonzero, the interval is simply shifted by that average. If $P(x,y)$ is
fluctuating as a function of $x$ and $y$ --- like it does for a disordered
medium --- the interval becomes a bit wider.) For electronic systems, $E$
gives the energy of the eigenstate, and Eq.~(\ref{Hdef})
hence describes a slab with a conduction band of width $8$. For photonic
systems, the real part of the eigenvalue gives the eigenfrequency. For both
systems, the imaginary part  $\gamma$ of the eigenvalue gives the decay rate of
the eigenstate. (Actually not $\gamma$ but rather
$\gamma/2$ is the decay rate but for the ease of
notation we will continue to refer to $\gamma$ simply as the decay rate.) 

$\kappa$ in Eq.~(\ref{Hdefkappa}) quantifies the strength of the
coupling between the slab and the outside.\cite{fyodorov:97a}
Using the units introduced
above, $\kappa$ has the value $\sin^2 k$ where $k$ is the wavevector
at the energy at which particles are injected respectively
emitted. This quantity is proportional to the velocity of the particle
perpendicular to the interface. In the centre of the band
$\sin k=1$ whereas at the edges $\sin k=0$.

If $\kappa$ is chosen to be constant (i.\,e. not to depend on energy) ideal
outcoupling can be described only for one specific value of the
energy. We will do this
since otherwise solving the Hamiltonian no longer is a standard eigenvalue
problem, and set $\kappa\equiv 1$, hence modelling ideal coupling at the centre
of the band.\footnote{It is not possible to have more than ideal coupling. For
$\kappa<1$ the loss rates are smaller than for $\kappa=1$, so this is easily
identified as ``sub-ideal''. For $\kappa>1$ the loss rates split into two
separate parts: Most become smaller, as for $\kappa<1$, while a few loss rates
become very large, thereby fulfilling the requirement that the average loss
rate has to be proportional to $\kappa$. We should note that this somewhat
counter-intuitive behaviour is also observed for chaotic
cavities.\protect\cite{fyodorov:97a}} Working at the centre of the conduction
bands offers the advantage that the width $N$ of the sample is identical to the
number $N$ of propagating modes, and thus allows the describe the largest number
of propagating modes for given size of the Hamiltonian (i.\,e. given numerical
work).
It is possible to include energy dependent
coupling terms\cite{terraneo:00a} but it should be stressed that a constant
$\kappa$ is more efficient and gives completely correct results as long as
only eigenvalues near the respective energy are considered. We set $\kappa=1$,
meaning ideal coupling at the centre of the conduction band.

It should be stressed that -- even though we are modelling a two-dimensional
system --- the results are valid for three-dimensional systems as long as 
$L\gtrsim N$.  A particle that is injected into the slab ergodically explores
the entire  cross-section of the sample before being emitted again,  and hence
loose its memory of which sites are ``connected'' by hopping terms. The sites
can then be re-arranged, e.\,g. in a three-dimensional structure. Only for very
short samples, $L\lesssim N$, this is not possible but for such samples already
applying the Anderson model (i.e. only allowing nearest-neighbour hopping) is
very questionable. The only ``real'' restriction that can limit the application
of our results to certain photonic three-dimensional systems is that particles
can leave the sample only at the front and at the back --- and not through the
``sides''.

\begin{figure}
\epsfig{file=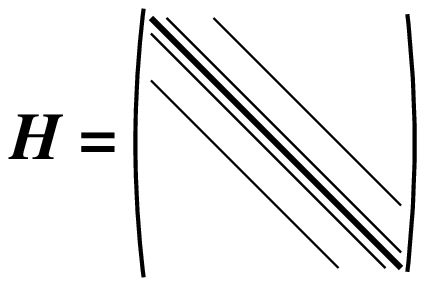,width=2cm}
\caption{\label{figmatrix}The Hamiltonian $\mathcal{H}$ has a band
structure. The thin lines contain matrix elements that are mostly
$1$, the diagonal is filled with complex numbers, and all other elements are
zero.
The entire matrix is symmetric but non-Hermitian (since there are
complex elements on the diagonal).}
\end{figure}

In the formulation of Eq.~(\ref{Hdef}) the matrix $\mathcal{H}$ has double
indices but these are easily removed by considering $\mathcal{H}_{n n'}$ with
$n=x+(y-1) N$. (It would not make sense to set $n=y+(x-1) L$ since usually $L\gg
N$, and we want to work with a band matrix that is as small as possible.) This
results is a matrix of the form as depicted in Fig.~\ref{figmatrix}. It is a
banded $L N \times L N$ matrix with band width $2 N + 1$. Also within the band
most elements are zero (since usually $N\gg 1$). The matrix is symmetric but
non-Hermitian as there are complex numbers on the diagonal.

The model~(\ref{Hdef}) has been widely used since an efficient
way to compute the transmission through such a slab is
known.\cite{baranger:91a} The method of recursive Greens functions allows to
compute the entire scattering matrix, hence all linear transport
properties, in a time of order $\mathcal{O}(L N^3)$ and with only minimal
storage requirements $\mathcal{O}(N^2)$. 
No explicit reference to the Hamiltonian
$\mathcal{H}$ is made, so that eigenvalues cannot be computed by this
method.

\section{Computing eigenvalues of symmetric complex non-Hermitian banded
matrices}
\label{secEigenSolver}

Since the Hamiltonian $\mathcal{H}$ from Eq.~(\ref{Hdef}) is both banded and
sparse
one might be tempted to use an
eigensolver for sparse matrices to compute the eigenvalues of Eq.~(\ref{Hdef}).
A sparse eigenvalue routine
needs to be  able to solve the equation
\begin{equation}
	( \mathcal{H} - \mu \eins ) \vec{x} = \vec{y}
	\label{inviteration}
\end{equation}
for the unknown vector $\vec{x}$ for arbitrary $\mu$ and $\vec{y}$. In particular, 
the eigensolver needs to
set $\mu$ close to an eigenvalue of $\mathcal{H}$ so that the matrix
$\mathcal{H} - \mu \eins$ is ill-conditioned. A numerical
solution of Eq.~(\ref{inviteration})
is then difficult and very expensive. Furthermore, only
one eigenvalue is found at a time, and control of which
eigenvalue the algorithm will converge to is difficult.
(Algorithms for sparse matrices
always use inverse iteration so that the corresponding eigenvector
will be returned without additional effort but the eigenvector is of no use
for us.) Using an algorithm for
banded matrices is thus the better alternative.

There exist a number of algorithms for real symmetric or complex
Hermitian band matrices. Both problems are characterised by real
eigenvalues, so that they are conceptually identical. Only one
algorithm for computing an eigenvalue (plus the corresponding
eigenvector) of a general complex band matrix has been
published.\cite{schrauf:91a} It uses inverse iteration, so it is of
limited use here.

We thus had to implement our own eigenvalue solver. 
The eigenrepresentation of $\mathcal{H}$ in terms of the diagonal 
matrix $\Lambda = \mathrm{diag}(\lambda_1,\ldots,\lambda_N)$ of the
eigenvalues $\lambda_i$ of A and the matrix $U$ of eigenvectors is
\begin{equation}
	\mathcal{H} = U \Lambda U^{-1} \;.
	\label{AewDecomp}
\end{equation}
We now observe that for symmetric, that includes \emph{complex} symmetric,
$\mathcal{H}$ it is always possible to chose eigenvectors such that $U
U^T = \eins$. If $U$ would be a real matrix, one would call $U$
orthogonal but since it is complex there is no special name  for
the property $U U^T = \eins$.

Algorithms for diagonalising a real symmetric matrix $A$ implicitly
decompose $A$ as
\begin{equation}
	A = Q D Q^T\;,\qquad Q Q^T = \eins
\end{equation}
with the matrix $D$ of eigenvalues of $A$. It is therefore possible to
adapt such an algorithm for our needs. Most algorithms for banded
matrices first reduce $A$ to tridiagonal form $A'$ by transformations
of the form $A'=Q' A Q'^T$, and we will adopt this strategy. 
(A matrix is called
tridiagonal if only the diagonal and its neighbouring elements are
nonzero. If $A$ would be real, the transformation $A\to A'$ would be
called a similarity transformation.) For a band matrix this is
possible in an efficient way since it is not necessary to compute (and
thus store) the full matrices $Q'$, and by annihilating the elements
of the matrix $A$ in a clever order, the band structure is kept intact in all
steps.\cite{kaufman:84a,kaufman:00a}

The reduction of the complex matrix $\mathcal{H}$ to tridiagonal shape
is done by straight-forward adaptation of this algorithm from real to
complex numbers, where care needs to be taken that the dot product
$\vec{x} \cdot \vec{y} = \sum_i x_i y_i$ is used and
not the dot product $\vec{x} \cdot \vec{y} = \sum_i \overline{x}_i
y_i$ normally used for complex vectors. (The overbar marks
the complex conjugate.)

To compute the eigenvalues of the tridiagonal matrix for
the real symmetric or complex Hermitian case, methods
that isolate eigenvalues in disjunct intervals are used (``divide and
conquer'' and similar methods\cite{golub:89a}). Such methods work for both
of these
cases as all
eigenvalues are real and can thus be ordered. This no longer is
possible here as the eigenvalues are complex. We therefore use
the QR respectively QL method.\cite{lapack:99} 

For an $K\times K$ banded matrix with band width $W$ the time needed
to compute the eigenvalues increases as $\mathcal{O}(K^2 W)$ whereas
the storage requirements increase as $\mathcal{O}(K W)$. In terms of
the dimensions $L$ and $N$ of the disordered slab, this means that
the time increases
as $\mathcal{O}(L^2 N^3)$ and the storage space as $\mathcal{O}(L
N^2)$. For given computational resources, both scalings impose an upper
limit on the system size that can feasible be treated.
For typical values of the ratio $L$ and $N$, and for ``realistic''
computer equipment, the time limit is reached somewhat earlier than
the memory limit.\footnote{On a modern computer a single
diagonalisation for a $L=700$, $N=70$ systems takes about $2$ days
and uses $256$ Mbytes of memory. While this memory requirement
frequently is no problem, the computing time usually is. Remember that
the task is to compute the distribution of the decay rates. Hence, many
matrices with different realisations of the random potential $P(x,y)$
have to be diagonalised --- not just a single matrix. However, the
restrictions imposed by time and memory are of the same order of magnitude.}

With respect to a similar algorithm for full matrices one wins a factor $L$
(usually of order several hundred) for both time and memory by using the banded
algorithm, thus allowing to treat system that could not be treated otherwise.
Still, the work presented in this paper is a big numerical challenge. To arrive
at the results, of the order of 100000 hours of cpu time on fast PC's were
 needed.

\section{Numerical simulations}

Disorder is modelled by assigning random values to $P(x,y)$. It is assumed that
those random numbers are uniformly distributed in the interval $[-w;w]$ so that
$w$ measures the amount of disorder. 

We only consider eigenvalues near the centre of the conduction band as the
assumption of ideal coupling is only valid there. For numerical reasons it is
essential that the centre of the conduction band is at $E=0$, i.e. one is
not allowed to add an offset to $P(x,y)$.\footnote{The algorithms will return
eigenvalues $z'$ that have a very small but finite deviation $|z-z'|$ from
their correct value $z$. Since we are primarily interested in the imaginary
part of the eigenvalue and want it to be as precise as possible the magnitude
of the real part has to be as small as possible.}

\begin{figure}[b!]
\epsfig{file=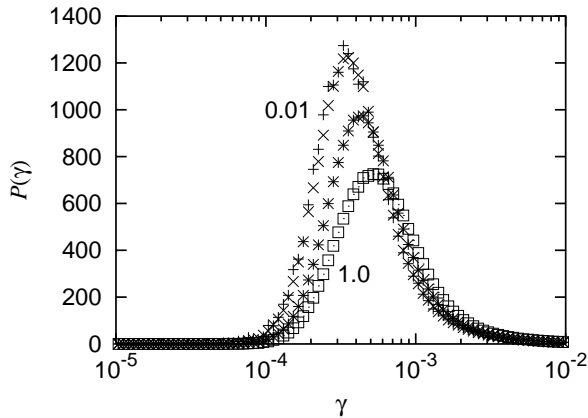,width=3.2in}
\caption{Probability distribution of the decay rates for given system parameters
as a function of the window size around the centre in which eigenvalues are
included in computing the probability distribution. The different dots mark the
distributions with $d=0.01$, $0.1$, $0.5$ and $1.0$.}
\label{windowsizefig}
\end{figure}

We hence chose a window $[-d;d]$ and only include eigenvalues in the further
analyses when their real part is inside that window. If the window is too
large, systematic errors are introduced while too small a window leads to bad
statistics. As can be seen from Fig.~\ref{windowsizefig}, for of $d=0.1$ the
distribution function already agrees with the distribution function for
$d=0.01$ but has much better statistics. $d=0.5$ and $d=1.0$ gives
significant systematic deviations. For this reason, all results presented
in this paper assume a window with $d=0.1$.

\begin{figure}
\epsfig{file=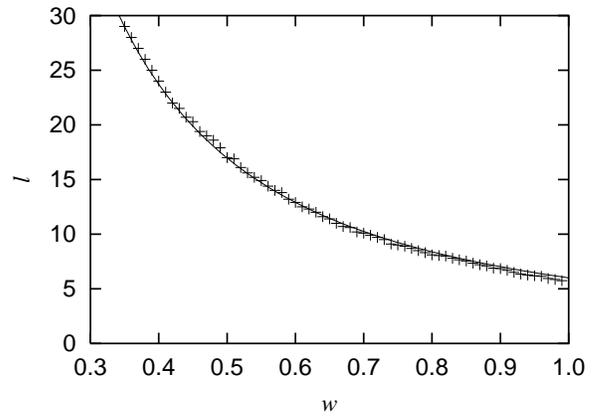,width=3.2in}
\caption{Mean free path $l$ as a function of disorder strength $w$ from a
numerical simulation (crosses) and from Eq.~(\protect\ref{freiwegeq}).}
\label{freiwegfig}
\end{figure}

The formulation of the model in Sec.~\ref{secModell} is in terms of generic
units. Contact with a microscopic model or an experiment is best made in terms
of the mean free path.
It can be computed from the length-dependence of the
transmission probability $T$ through the sample.
In the diffusive regime, $l\lesssim L\ll N l$, it
is given by~\cite{beenakker:97a}
\begin{equation}
	\frac{1}{T} = 1 + \frac{L}{l} \;.
\end{equation}
The mean free path can be computed by fitting $T(L)$ to this functional form.

The transmission probability has been computed using the method of recursive
Green's functions\cite{baranger:91a} for variable disorder strength $w$. As
Fig.~\ref{freiwegfig} shows, the numerically computed
mean free path $l$ is for the range of $w$ in question in very good
approximation given by
\begin{equation}
	l = \frac{6}{w^{3/2}} \;.
	\label{freiwegeq}
\end{equation}
(Computed for each value of $w$ from $50$ samples with $L=2,4,\ldots,98,100$ 
and $N=50$.)
In the following, we will no longer make explicit reference to $w$ but rather
give the more intuitive mean-free path $l$.

\section{Diffusive regime}
\label{secDiffusive}

For a sample length $L$ with $l\lesssim L \ll N l$ the sample is said to be in
the diffusive regime. It is immediately obvious that the diffusive regime can
only be observed in sufficiently wide samples, $N\gg 1$.

For chaotic cavities with broken time-reversal symmetry an analytical
result for the decay rate distribution has been given by Fyodorov and 
Sommers.\cite{fyodorov:97a}
We start from their result and rescale it,
\begin{equation}
	\mathcal{P}(y) = \frac{1}{y^2 M!} \int_0^{M y} x^M e^{-x} d x
		= \frac{1}{y^2} \left[ 1 - e^{-M y} \sum_{k=0}^M \frac{M^k}{k!}
		y^k \right] \;.
	\label{PyAnsatz}
\end{equation}
$\mathcal{P}(y)$ is normalised to one and in our scaling is for all $M>1$ 
peaked near a value of $y$ of order $1$. In the original formulation for a 
chaotic cavity, $M$ is the number of modes propagating through the opening of 
the cavity.

\begin{figure}
\epsfig{file=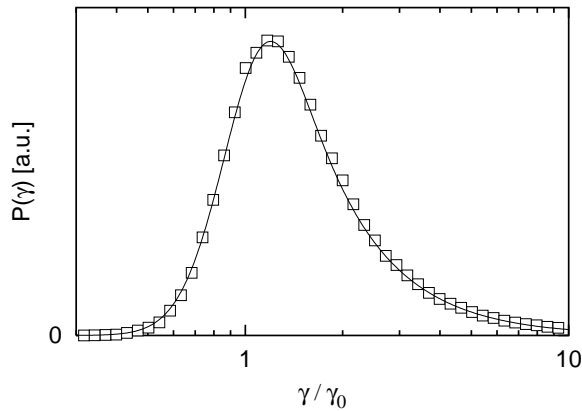,width=3.2in}
\caption{The numerically computed probability distribution $P(g)$ for
$L=175$, $N=50$, $l=12.9$ and comparison with Eq.~(\ref{PyAnsatz}) with $M=16$.}
\label{PyGueltigFig}
\end{figure}

In the following we will argue that the decay rate distribution $P(g)$ can be
written in the form~(\ref{PyAnsatz}) as
\begin{equation}
	P(\gamma)=\frac{1}{\gamma_0}
		\mathcal{P}\bigl(\frac{\gamma}{\gamma_0}\bigr)
\end{equation}
with some
scaling factor $\gamma_0$ and some effective number of modes $M\ne N$. In
Fig.~\ref{PyGueltigFig} a comparison between the analytical suggestion and a
simulation is given, and the agreement is striking. The horizontal axis has
been plotted logarithmically since this results in both the differences between
the $P(y)$ for different $N$ becoming easier to recognise and in giving a more
prominent place to the small-$\gamma$ tail of $P(\gamma)$. In most applications,
including the random laser discussed later in this paper, one is much more
interested in small $\gamma$ than in large $\gamma$.

\begin{figure}[b!]
\epsfig{file=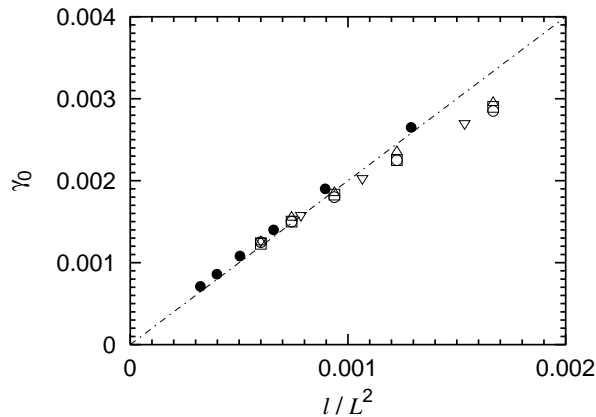,width=3.2in}
\caption{Scaling factor $\gamma_0$ as a function of $l$ and
$L$ for $l=24$ and $N=10,
20, 50, 100$ (open symbols) and $l=12.9$ and $N=50$ (solid circles). The line
marks the prediction from Eq.~\protect\ref{g0Gleichung}.}
\label{g0Grafik}
\end{figure}

The results of the simulations are fitted ``by eye'' against the functional
form~(\ref{PyAnsatz}), resulting in one pair of values for $\gamma_0$ and $M$
for 
each
set of parameters. 
Especially at very small $\gamma$, there are sometimes numerical errors that
introduce artifacts into the numerical histogram so that using 
an automatic fitting
algorithm is not feasible.
(Usually we computed 500--1500 realisations for each
parameter set.)
From our simulations, we find that the scaling factor $\gamma_0$
only depends on the
length $L$ of the sample and its mean free path $l$ but not on its width $N$,
and seems to be given by
\begin{equation}
	\gamma_0 = \frac{2 l}{L^2} \;.
	\label{g0Gleichung}
\end{equation}
As Fig.~\ref{g0Grafik} shows, the agreement between the result of
the numerical simulations and Eq.~(\ref{g0Gleichung}) is
good, and all major deviations are for small $L$ where universal scaling is
expected to be worse than for larger $L$. The model equations set the
speed of propagation to $1$ but it is obvious that for some other choice for the
propagation speed $c$ one has to change Eq.~(\ref{g0Gleichung}) to 
$\gamma_0 = 2 c l / L^2$.

\begin{figure}[t!]
\epsfig{file=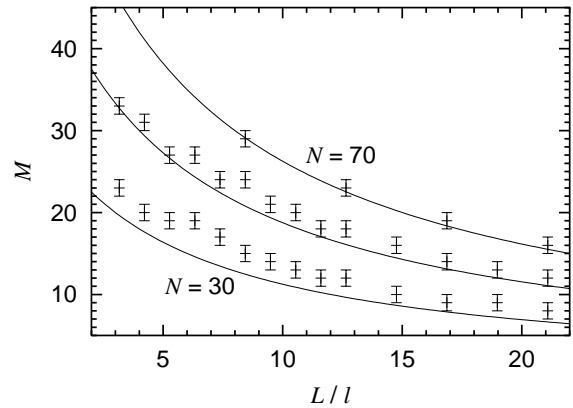,width=3.2in}
\caption{Effective number $M$ of degrees of freedom as a function of the length
$L$ of the sample for a sample with $N=30$, $N=50$ respectively $N=70$
propagating modes. The line marks the prediction from
Eq.~\protect\ref{Mgleichung}, the points are from numerical simulations.
The size of  the ``errorbars'' does not indicate some estimated error interval
but simply marks the computed value $\pm 1$.}
\label{Mgrafik}
\end{figure}

While the determination of $\gamma_0$ is very precise, there is a somewhat
larger error involved in determining $M$ by fitting the analytical form to the
results of numerical simulations. First, we only fitted against integer $M$,
though in principle a generalisation of Eq.~(\ref{PyAnsatz}) to noninteger $M$
is possible, see Eq.~(\ref{PvertGamma}). Secondly, if $M \gtrsim 25$,
the  difference
between $P(y)$ for $M$ and for $M+1$ becomes too small to tell with certainty
which
of these two values describes the numerical result better. Thirdly and finally,
even with 500-1500 samples for each set of parameter values, there are still
some fluctuations in the numerically computed histogram for the decay rate
distribution that in some cases make the decision on the right $M$ a bit
difficult. Considering all of this, one should allow for an error of $1$
for $M$, and even of $2$ for $M\gtrsim 25$.

We have computed $M$ for a series of samples with increasing length
for three different widths $N$. 
As Fig.~\ref{Mgrafik} shows, the effective number $M$ of modes is
well approximated by
\begin{equation}
        M = \frac{N}{1 + L / (6 l)} \;.
        \label{Mgleichung}
\end{equation} 
The agreement between this suggested analytical form and the numerical
simulation becomes better as the width $N$ of the sample is increased. From the
simulations it is obvious that the functional form Eq.~(\ref{Mgleichung}) 
is correct but there still is the (small) possibility that the factor $6$ might
need to be replaced by a slightly smaller value. To answer this question with
certainty, we would need to increase both $L$ and $N$ significantly.
Unfortunately, such simulations are outside the present time and memory
constraints.

Equations~(\ref{PyAnsatz}--\ref{Mgleichung}) give a good description
of the decay rate distribution of a disordered slab in the diffusive regime,
provided the slab is sufficiently wide. Since the transversal length scales are
set by microscopic quantities (wave length of the light for optical systems,
Fermi wave length for electronic systems), all macroscopic objects are
``wide''.

\renewcommand{\bottomfraction}{0.9}
\renewcommand{\textfraction}{0.1}
\renewcommand{\dblfloatpagefraction}{0.9}

\begin{figure}[t]
\epsfig{file=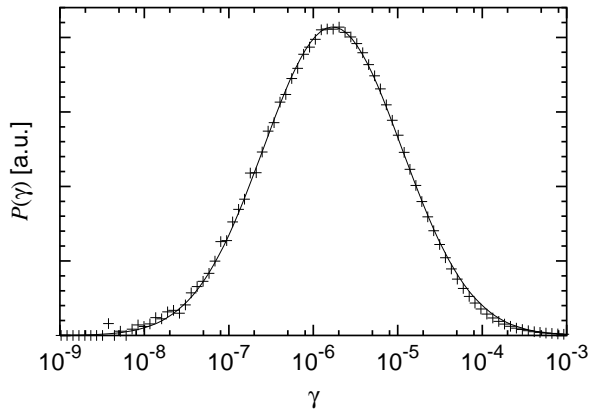,width=3.2in}
\caption{Numerically computed distribution of the decay rate $\gamma$ for a
sample in the localised regime ($L=71.55~l$, $N=15$) and comparison with the
log-normal distribution~(\protect\ref{pLokalAnsatz}).}
\label{logbeispiel}
\end{figure}

\begin{figure*}[tb!]
\epsfig{file=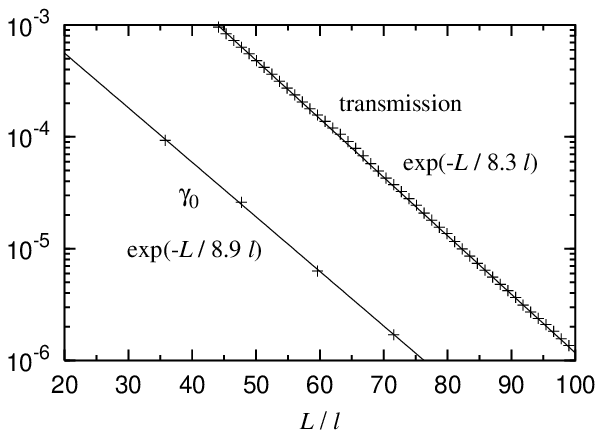,width=5.9cm}
\hfill
\epsfig{file=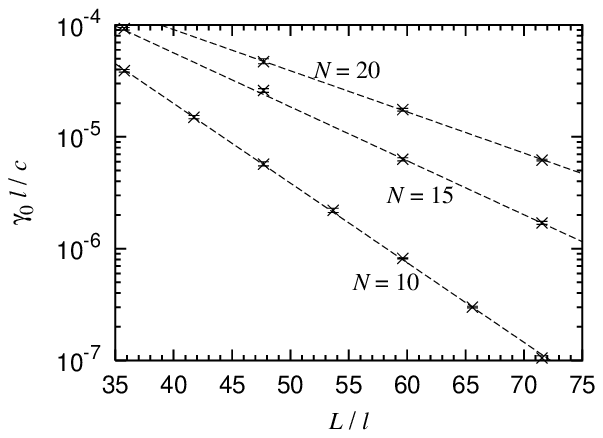,width=5.9cm}
\hfill
\epsfig{file=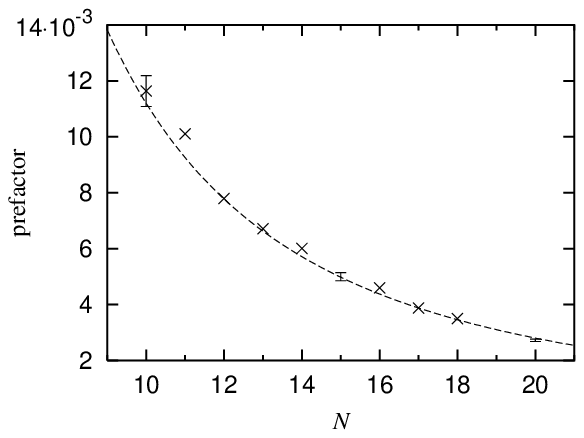,width=5.9cm}
\caption{Left: Transmission through the sample and position $\gamma_0$ of
the maximum of the log-normal distribution as a function of the length of the
sample (for $N=15$). There is a small
but finite difference between the length scales for
both quantities. Centre: Position $\gamma_0$ of
the maximum of the log-normal distribution as a function of length
for samples of different width $N$. 
Right: Prefactor in front of the exponential for
$L=71.55~l$ for different $N$ [cf. Eqs.~(\protect\ref{ylocposprop})
and~(\protect\ref{ylocpos})].
The dashed line marks the curve $1.12/N^2$.
}
\label{figlocmax}
\end{figure*}

\section{Localised regime}
\label{secLocalised}

If the length $L$ of a disorder medium is increased, the phenomenon of
localisation sets in once $L\gtrsim N l$ (see Ref.~\onlinecite{beenakker:97a}
for a review). In the localised regime the
probability of transmission $T$ through the sample is reduced significantly and
decays exponentially with the length $L$ of the sample. The length scale
$\xi$ is called the localisation length, and can be
computed from an ensemble of disordered slabs 
by computing
the average of the logarithm of the transmission as a function of the length of
the samples, hence
\begin{equation}
	-L/\xi = \langle \ln T(L) \rangle \;.
\end{equation}
One should note that this is not identical to fitting the transmission
to $\langle T(L)
\rangle \propto \exp( -L/\xi )$ since the large sample-to-sample fluctuations of
$T(L)$ in the localised regime would give a value for $\xi$ that is off by a
factor
$4$. 
The localisation length can also be computed analytically
from the mean-free path using the DPMK equation, with the
result\cite{efetov:83b,efetov:83c,dorokhov:82a,dorokhov:82b,dorokhov:83a,dorokhov:83b}
\begin{equation}
	\xi = \frac{N+1}{2} l \;,
	\label{XiVorhersage}
\end{equation}
and agrees well with our numerical results.

It is generally accepted that the distribution of the decay rates $\gamma$ (at
least for small $\gamma$) in the localised regime is log-normal, i.\,e., $\ln
\gamma$ is distributed according to a Gaussian distribution. Recent interest
has rather been in the large-$\gamma$ tail which was shown to follow a
power-law.\cite{titov:00a,terrano:00a} In a log-normal distribution, most of
the weight lies in the right tail, so those papers give a sufficient
description for most of the eigenmodes. In the context of applications to
random lasers we are, however, interested in the small decay rate tail, hence
in the log-normal distribution. 

To our knowledge, there is only a single paper by
Titov and Fyodorov that gives explicit expressions for the
parameters of that log-normal distribution.\cite{titov:00a}
However, their analytical results are for a somewhat different system
so it is difficult to tell whether they agree or disagree 
with our findings. We will
return to this aspect at the end of this section.
First, we want to present the results of our
numerical simulations.

Using the log-normal ansatz, the distribution of the decay rates $\gamma$ is
\begin{equation}
	P(\gamma) = b \exp\left(-\frac{(\log \gamma - \log
	\gamma_0)^2}{\sigma^2} \right) \;.
	\label{pLokalAnsatz}
\end{equation}
The numerically computed histograms indeed follow this form, see
Fig.~\ref{logbeispiel}, except for the large-$\gamma$ tail ---
as already mentioned above but this deviation is only seen in a log-log plot.

Fig.~(\ref{figlocmax}) shows in the left 
the numerically computed $\gamma_0$ as a function
of $L$ for $N=15$. Also displayed is the localisation length $\xi$
computed numerically from the transmission, being in good agreement with
the analytical
prediction~(\ref{XiVorhersage}). The quality of the data is good enough to say
with confidence that $\gamma_0$ decays exponentially with a length scale that
is somewhat larger than $\xi$. Fig.~(\ref{figlocmax}) shows in the right
the value of
$\gamma_0$ also for two other values of $N$, and all three cases are
well-described by introducing a numerical fitting factor $a$,
\begin{equation}
	\gamma_0 \propto \exp\left(-\frac{L}{a \xi}\right) \quad
	\mathrm{with}\quad a=1.12 \;.
	\label{ylocposprop}
\end{equation}
It is known that working at the centre of the conduction band when
in the localised regime can introduce certain artefacts, especially in
analytical approaches.
Among other, the localisation length at the band
centre can differ by approximately $10\,\%$ from the value outside the 
centre.\cite{kappus:81a} We have
defined $\xi$ based on the transmission through the sample (at an energy
corresponding to the band centre), and in transmission resonances at all
energies can contribute. 
A numerical prefactor $a$ that differs by about $10\,\%$
from $1$ thus does not come as a complete surprise.

We still need to compute the proportionality factor appearing in
Eq.~(\ref{ylocposprop}). For this purpose we need to plot the ratio of the
numerically computed $\gamma_0$ and the right-hand side of
Eq.~(\ref{ylocposprop}) for different values of $N$.
We did this for $L=71.55~l$. Since this is a
very expensive operation, we have computed a large number of samples 
only for $N=10,15,20$ 
so that
their statistics is better than for the other values of $N$. An estimate of the
error for these ``better'' data points has been included in the figure. This
allows us to conclude that
\begin{equation}
	\gamma_0 = \frac{a}{N^2} \exp\left(-\frac{L}{a \xi}\right) \quad
	\mathrm{with}\quad a=1.12 \;.
	\label{ylocpos}
\end{equation}
It should be noted that this equation contains two numerical coefficients, and
there is no obvious reason why they should be identical. Still, we find
that they both are approximately $a=1.12$.

Re-introducing ``physical units'' into Eq.~(\ref{ylocpos}) is a bit more
difficult than it was for Eq.~(\ref{g0Gleichung}) where it was obvious that one
simply has to multiply by the velocity of propagation $c$. Here one has to
multiply by $c/\Delta$ where $\Delta$ is the perpendicular grid spacing. Due to
the assumption of one propagating mode per (lateral) grid point made in
Sec.~\ref{secModell}, $\Delta$ is not arbitrary but has a well-defined
physical meaning. For the electronic case, $\Delta=\pi/k_{\mathrm{F}}$ with
$k_{\mathrm{F}}$ the wave vector at the Fermi level, and for the photonic case
$\Delta=2\lambda/\pi$ with $\lambda$ the wave length of the light (hence
$c/\Delta = 1/(4\nu)$).

\begin{figure}
\epsfig{file=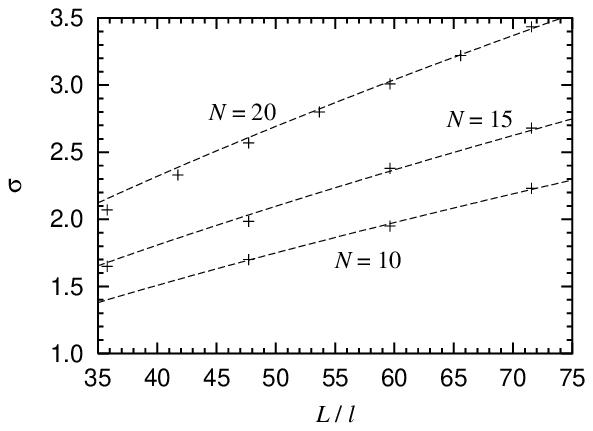,width=3.2in} \\
\epsfig{file=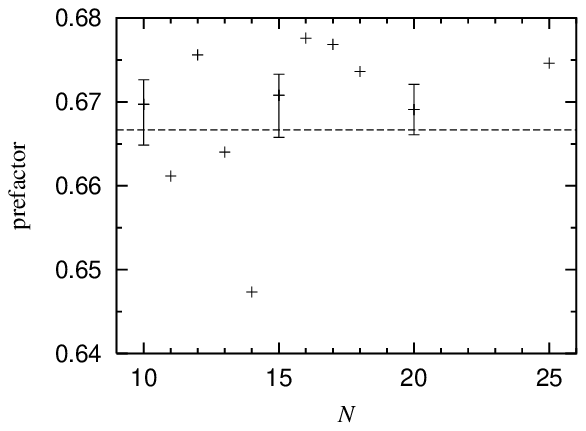,width=3.2in}
\caption{Top: Comparison of the numerically computed $\sigma$ with the
prediction from Eq.~(\protect\ref{eqsigma}). Bottom: Comparison of the
numerically
computed prefactor [from dividing the numerical $\sigma$ by $(L/a\xi)^{2/3}$,
cf. Eq.~(\protect\ref{eqsigma})] 
with the prediction $2/3$ for samples 
of
$L=71.55~l$. For some values of $N$ a higher number of samples has been
computed, so that the data quality is high enough to also compute an error
estimate. This has been indicated in the figure.}
\label{figsigma}
\end{figure}

Determining the width $\sigma$ of the distribution is more difficult since we
can only use the left wing of the distribution --- the right  wing eventually
turns into a power-law tail and thus no longer follows a log-normal distribution.
Once again, we have accumulated more data for $N=10,15,20$ so that some
indication of the error is possible for those three data points.

From our data, we propose the formula
\begin{equation}
	\sigma = \frac{2}{3} \left(\frac{L}{a \xi}\right)^{2/3} \;,
	\label{eqsigma}
\end{equation}
where $a=1.12$ has the same value as in Eq.~(\ref{ylocpos}). As
Fig.~\ref{figsigma} shows, there clearly is no
disagreement between the numerical data and Eq.~(\ref{eqsigma}). However,
please remember that the $\frac{2}{3}$ should be thought of as a fitting factor 
that might not be exactly $2/3$ but perhaps rather $0.67$ or some other
numerical factor.

Since the distribution is log-normal only for not too large $\gamma$ (remember
the power-law tail for large $\gamma$) the normalisation
is nontrivial [$P(\gamma)$ is not normalised to $1$ any longer!] and
cannot be computed from $\gamma_0$ and $\sigma$. The constant $a$ in
Eq.~(\ref{pLokalAnsatz}) is directly equal to the height of the peak of the
numerically computed $P(\gamma)$. Since the total area underneath the numerically
computed $P(y)$ (and hence its normalisation) is dominated by the large-$\gamma$
tail, $a$ has a relatively large error.
Taking all the available data, the most likely value is
\begin{equation}
	b=\frac{1}{N^2} \exp\left(\frac{L}{a\xi}\right) \;.
\end{equation}
This value has been determined from a large number of simulations that for space
reasons cannot be presented here. 
Unfortunately the quality of the data is not good
enough to decide whether an additional prefactor $a=1.12$ should appear.

At the present it is not possible to tell whether our results agree with the
ones put forward by Titov and
Fyodorov.\cite{titov:00a} In particular, they arrive at
\begin{equation}
	\gamma_0 \propto \exp\left(-\frac{3 L'}{\xi}\right) \;,
	\label{EQmisha}
\end{equation}
whereas our finding~(\ref{ylocpos}) was $\gamma_0 \propto \exp(-L/1.12\xi)$.
There are two obvious differences between the model used by them
and the model employed by us. First, for numerical reasons we work at
the centre of the conduction band while they work near (but 
sufficiently far away from) the band
edges. This might explain the factor $a=1.12$ that we have to introduce.
Secondly and probably more importantly, they consider a system of length
$L'$ that is closed at one end whereas our systems have length $L$ and are open
at both ends. It is obvious that a half-closed system of length $L'$ corresponds
to an open system of length $L>L'$. Eq.~(\ref{EQmisha}) suggests that those two
systems could be mapped into each other by setting $L\approx 3 L'$ but there is
no further evidence to support this claim.

\section{Lasing threshold of a random laser}
\label{secThreshold}

A random laser is a laser where the necessary feedback is not due to mirrors at
the ends of the laser but due to random scattering inside the
medium.\cite{wiersma:95a,wiersma:97a,beenakker:98b}
We model the random laser as a disordered slab containing a dye that is able to
amplify the
radiation in a certain frequency interval with rate $1/\tau_{\mathrm{a}}$. The
lasing threshold is the amplification rate at which the intensity of
the emitted radiation
diverges in a linear model. If saturation effects are included, the emitted
intensity increases abruptly but finitely at crossing the lasing threshold.

The lasing threshold is given by the value of the smallest decay rate of all
eigenmodes in the amplification window.\cite{misirpashaev:98a} (Remember that
$\gamma$ actually is twice the decay rate. On the other hand, also
$1/\tau_{\mathrm{a}}$ enters the relevant formulae only with a prefactor $1/2$.
$\gamma$ thus indeed gives the necessary amplification rate 
$1/\tau_{\mathrm{a}}$.) This is easily
understood since this simply means that in the mode with the smallest decay
rate the photons are
created faster by amplification than they can leave the sample (=decay). 
It, however, also follows from a complete quantum mechanical
analysis.\cite{schomerus:00a,patra:99a} 

%
%
%

The distribution of the decay rate has been computed in this paper.
A
certain number $K$ of modes will be in the frequency window where
amplification is possible. The lasing threshold is given by the smallest $\gamma$
of these $K$ modes. In a simple picture that is valid once $K\gg 1$ we can
assume that the $K$ different $\gamma$'s are distributed independently
according to $P(\gamma)$.\cite{misirpashaev:98a} The distribution
$\tilde{P}(\gamma)$ of the
smallest mode and hence of the lasing threshold then becomes
\begin{equation}
	\tilde{P}(\gamma) = K P(\gamma)\left[ 1-\int_0^{\gamma}
		 P(\gamma') d\gamma' \right]^{K-1} \;.
	\label{eqtreshold1}
\end{equation}

For $K\ne 1$, the distribution $\tilde{P}(\gamma)$ of the lasing threshold is 
not longer identical to the distribution $P(\gamma)$ of the decay rate of each
individual mode. In particular, not only the precise form of these two
distribution will be different, but also the ``typical'' value of the lasing
threshold can be different from the ``typical'' decay rate $\gamma_0$.
Interestingly, for chaotic cavities in the diffusive regime
it was found that the latter two quantities
differ only insignificantly\cite{frahm:00a,schomerus:00a} which might seem
counter-intuitive. A slab geometry is more ``complicated'' in that the scaling
$K\propto N$ ``tries'' to lower the lasing threshold with increasing $N$.

For $K\gg 1$ the distribution $\tilde{P}(\gamma)$ 
is sharply peaked around its maximum. The
position
$\gamma_{\mathrm{m}}$ of the maximum is given by
the solution of the equation
$d \tilde{P}(y_{\mathrm{m}})/d\gamma_{\mathrm{m}}=0$, hence
\begin{equation}
	0 = \frac{d P(\gamma_{\mathrm{m}})}{d\gamma_{\mathrm{m}}} 
		\left[ 1-\int_0^{\gamma_{\mathrm{m}}} P(\gamma') d\gamma' 
		\right]
		- (K-1) [ P (\gamma_{\mathrm{m}})]^2 \;.
	\label{eqtreshold2}
\end{equation}
While Eq.~(\ref{eqtreshold1}) is difficult to compute numerically due to the
large exponent $K-1\gg 1$, in Eq.~(\ref{eqtreshold2}) this exponent no longer
appears.

Eq.~(\ref{eqtreshold2}) depends on $P(\gamma)$ which in turn depends on the
dimensions $L$ and $N$ of the system. In assuming that the number of propagating
modes is equal to the width $N$ of the sample we already have made the
assumption that the width (and hence also the length) 
is measured in units of $\lambda/2$. (The ``$2$'' accounts for polarisation.)
The total number of modes in the sample thus is $L N$. We assume that a fraction
$f$ of them is inside the amplification window of the dye, hence $K=f N L$. For 
simplicity we neglect complications
as the shape of the mode profile. (It is easily incorporated into the numerics
and we refrain from doing this just to prevent having to introduce even more
parameters.) $f$ depends only on the chemical properties of the dye and not
on the dimensions of the sample.

In the following we will show how to compute the most likely lasing threshold for
samples in both the diffusive and in the localised regime.

\subsection{Lasing threshold in the diffusive regime}
\label{secThresholdDiffusive}

The change of the lasing threshold with increasing system size is influenced
by a subtle interplay between $L$ and $N$ in determining the distribution
$P(\gamma)$ and in determining the number $K=f N L$ of total modes.

If $K\gg 1$ the lasing mode has a
decay rate in the low-$\gamma$ tail of $P(\gamma)$ (i.e. $\gamma<\gamma_0$ or
$y<1$).
The weight of this tail is 
\begin{equation}
	\int_0^1 P(y) d y = \frac{M^{M-1}}{(M-1)!} e^{-M} \;,
	\label{P0bis1}
\end{equation}
and goes to zero as $M$ becomes larger.
For $M\to\infty$ the tail disappears completely, as is already obvious from
the asymptotic form of the distribution,
\begin{equation}
	P_{M\to\infty}(y)=\left\{ \begin{aligned}
		& 0 & (y<1) \\
		& 1/y^2 & (y\ge 1)
		\end{aligned} \right.
\end{equation}
With increasing $N$ and hence increasing $M$, the probability that a given mode
has a small $y$ thus decreases rapidly. On the other hand, we are interested in
the smallest decay rate out of
 $K$ modes, and $K$ increases linearly with $N$. This
are two counter-acting effects, and it is not obvious which of these two is
stronger. 

The effect of an increase of the system size $L$, on the contrary, is obvious.
First, the average decay rate $\gamma_0$ decreases according to
Eq.~(\ref{g0Gleichung}). Secondly, $M$ decreases from Eq.~(\ref{Mgleichung}),
leading to even smaller values for $\gamma$ of the lasing mode.

There have been some analytical attempts to compute the lasing threshold for a
chaotic cavity~\cite{frahm:00a,schomerus:00a}.
For large $M$, the small-$y$ tail of Eq.~(\ref{PyAnsatz}) was approximated by
\begin{equation}
        P(y) \approx \frac{1}{2 M} 
                \left[ 1 + \mathrm{erf}\bigl( \sqrt{M/2} [y-1]
                \bigr) \right] \;.
	\label{Pyapproximation}
\end{equation}
This allows to arrive at scaling laws of the lasing threshold for variable $M$
at fixed $K$. Unfortunately, the difference between two counter-acting effects
of an increase in $N$ are so small that Eq.~(\ref{Pyapproximation})
is a bit too crude for our needs.

We thus have to revert to a numerical procedure. Eq.~(\ref{PyAnsatz}) can be
rewritten using the incomplete Gamma function
\begin{equation}
	\Gamma(a,x)=\int_x^{\infty} t^{n-1} e^{-t} d t \;.
\end{equation}
$\Gamma(a,0)$ reduces to the well-known Gamma function $\Gamma(a)$. For
numerical reasons it is advisable to introduce the regularised Gamma function
$Q(a,x)=\Gamma(a,x)/\Gamma(a)$. Fast numerical algorithms to compute $Q(a,x)$
exist. [Please note that in the literature the
definitions of the regularised Gamma function sometimes 
disagree in that our $Q(a,x)$
is denoted as $1-Q(a,x)$.] Now we can express Eq.~(\ref{PyAnsatz}) and
its derivative and integral as
\begin{subequations}
\label{PvertGamma}
\begin{align}
	P(y) &= \frac{1}{y^2} \bigl[1 - Q(M+1, M y)\bigr] \;, \\
	\frac{d P(y)}{d y} &= \frac{(M y)^M}{y^2 \Gamma(M)} e^{-M y}
		 - \frac{2}{y^3}\bigl[1- Q(M+1, M y)\bigr] \;, \\
	\int_0^y P(y') d y' &= \frac{1}{y} \bigl[Q(M+1,M y)-1\bigr] + 1 
		- Q(M,M y) \;,
\end{align}
\end{subequations}
so that Eq.~(\ref{eqtreshold2}) can be evaluated efficiently.

\subsection{Lasing threshold in the localised regime}
\label{secThresholdLocalised}

From Eq.~(\ref{pLokalAnsatz}) we directly arrive at
\begin{subequations}
\begin{align}
	\frac{d P(\gamma)}{d \gamma} &=
		-2 b \frac{\ln\gamma-\ln\gamma_0}{\gamma\sigma^2}
			\exp\left[
			-\frac{(\ln\gamma-\ln\gamma0)^2}{\sigma^2}\right]
		\;, \\
	\int_0^{\gamma} P(\gamma') d \gamma' &= 
		\frac{b \sqrt{\pi} \sigma \gamma_0}{2}
		 e^{\sigma^2/4} \left[1+\mathrm{erf}\left(
	 	\frac{2 \ln( \gamma / \gamma_0)-\sigma^2}{2\sigma}\right) 
		\right] \;.
\end{align}
\end{subequations}
A further simplification is not possible, and we did not manage to find 
suitable
approximations. Also for the localised regime we thus are restricted to a
numerical evaluation.

\subsection{Numerical results}

\begin{figure*}[t!]
\epsfig{file=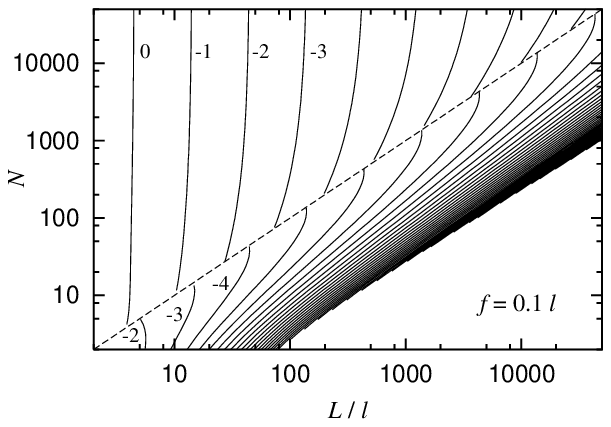,width=5.9cm}
\hfill
\epsfig{file=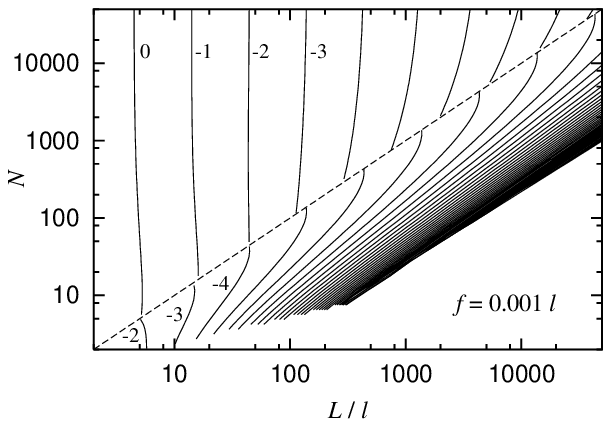,width=5.9cm}
\hfill
\epsfig{file=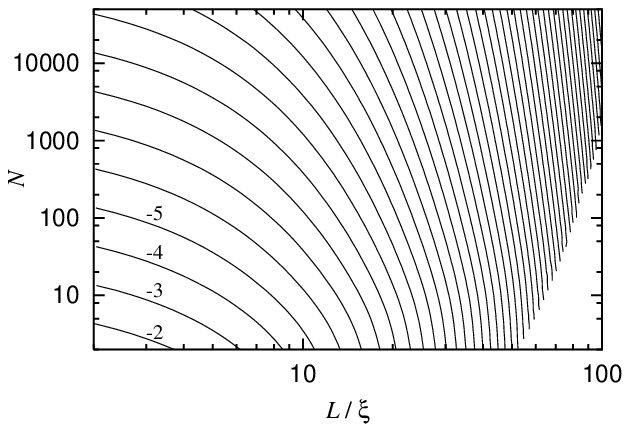,width=5.9cm}
\caption{Most likely lasing threshold as a function of the length $L/l$ and the
width $N$ of the sample. For $L/l \gtrsim N$ (the lower right part of the
diagram) the sample is localised, for $L/l \lesssim N$ (the upper left part) the
sample is diffusive. The diagonal line marks the division, and the results
close to that line should thus be viewed with caution. 
The left diagram depicts
the results for $f=0.1$, the central diagram for $f=0.001$. The right diagram
again depicts the results for the localised regime with $f=0.1$ but the
horizontal axis has been rescaled to $L/\xi$ instead of $L/l$. The numbers $x$
at the contour lines mean $10^x c/l$ in the diffusive regime, and $10^x
c/\Delta$ in the localised regime.}
\label{figLaserschwelle}
\end{figure*}

The lasing threshold is computed numerically from Eq.~(\ref{eqtreshold2}),
using the formulae from Secs.~\ref{secThresholdDiffusive}
and~\ref{secThresholdLocalised}. Into the formulae presented there, we have to
insert the correct dependence of the $\gamma_0$, $M$, $\sigma$, etc. on $L$ and
$N$ that was presented earlier in this paper. Despite this complication the
numerical calculation is straight forward as Eq.~(\ref{eqtreshold2})
possesses a single root only. Since this root has a change of sign, it is easily
found numerically.

Fig.~\ref{figLaserschwelle} shows the results for both the diffusive and the
localised regimes, for both $f=0.1~l$ and $f=0.001~l$. (The mean-free path
appears as a factor since the figure is in units $L/l$ and not $L$.) The
formulae found in this paper are valid deep in the diffusive regime
respectively deep in the localised regime.  Near the cross over, hence near the
line $L\approx N l$, this condition is not fulfilled. The numerical values 
near the  diagonal line 
in
Fig.~\ref{figLaserschwelle}  should thus be viewed with
caution.

As can be seen from the figure, in the diffusive regime with $N\gg L/l$
the lasing threshold becomes almost
independent of the width $N$ of the sample (for sufficiently large $N$), 
and the most likely value of the
lasing threshold is about
\begin{equation}
	\gamma_{\mathrm{m}} \approx \frac{2 c l}{L^2} \;,
	\label{gammaMdiffuse}
\end{equation}
hence the value given by Eq.~(\ref{g0Gleichung}). This means that even though
$K\gg 1$, $P(y)$ for $y<1$ is already so small that it dominates over the large
value of $K$. Differences between this simple approximation and the precise
numerical result appear for finite $N$, with the size of this difference
depending on $f$. However,
for designing experiments it is obvious from the results presented here
that the only feasible way to lower the
lasing threshold of a random laser in the diffusive regime is increasing its
length, not modifying its width.

As Fig.~\ref{figLaserschwelle} shows, also in the localised regime
there is only a small dependence on $f$.
This means that in a log-normal
distribution the weight of the left tail is so small that unless $K$ is
exponentially large $\gamma_{\mathrm{m}}$ cannot become much smaller than the
position $\gamma_0$ of the peak of the distribution. The difference to the
diffusive regime is that the lasing threshold can be lowered efficiently 
not only by increasing the length but also 
decreasing the width $N$ and hence driving the system farther into
localisation. 

It is no surprise that samples in the localised regime generally have a lower
lasing threshold than samples in the diffusive regime. We have shown that also
the diffusive samples can have an ``acceptably small'' lasing threshold as it 
is trivial to make them very long (since there is no need to care much about
their width). For both the diffusive and the localised regime, the typical decay
rates of a single mode are comparable to the lasing threshold.

\section{Conclusions}
\label{secConclusions}

We have numerically computed the distributions of the residues (or decay rates)
of a disordered slab.
The slab has length $L$, mean free path $l$, width respectively cross-sectional
area $N$ ($N$ is given as
number of propagating channels) and velocity of propagation $c$. We were able to
``guess'' simple analytical formulae that are able to describe the numerical
results well.

For a sample in the diffusive regime ($L\lesssim N l$) we found in
Eqs.~(\ref{PyAnsatz}--\ref{Mgleichung})
\begin{subequations}
\label{ergebnis1}
\begin{align}
        P(\gamma)&=\frac{L^2}{2 l c}
                \mathcal{\mathcal{P}}\bigl(\frac{\gamma L^2}{2 l c}\bigr) \;, \\
	\mathcal{P}(y) &= \frac{1}{y^2} \Bigl[1 - 
	\frac{\Gamma(1+\frac{N}{1 + L/6 l},\frac{N y}{1 + L/6 l})}{
		\Gamma(1+\frac{N}{1 + L/6 l})} \Bigr] \;,
\end{align}
\end{subequations}
where $\Gamma(a,x)$ is the incomplete Gamma function. The agreement between the
numerical results and the proposed formulae is good, and there is
the possibility that Eq.~(\ref{ergebnis1})
 might become exact in the limit $L/l\gg N \gg 1$.
However, there is only numerical and no analytical evidence 
to back this claim.

For a sample in the localised regime ($L \gtrsim N l$) with localisation
length $\xi=(N+1)l/2$ we found in Sec.~\ref{secLocalised}
\begin{gather}
        P(\gamma) = \frac{1}{N^2} \exp\left(\frac{L}{a\xi}
		-\frac{(\log \gamma - \log
        \gamma_0)^2}{\sigma^2} \right) \;, \quad a=1.12 \;, \nonumber\\
        \gamma_0 = \frac{a}{N^2} \exp\left(-\frac{L}{a \xi}\right) \;,\quad
	        \sigma = \frac{2}{3} \left(\frac{L}{a \xi}\right)^{2/3} \;.
	\label{ergebnis2}
\end{gather}
The quality of the simulations results in the localised regime is somewhat less
than in the diffusive regime. For this reason, Eq.~(\ref{ergebnis2}) should be
understood as an approximate fit only, and it very probably differs from the
exact relation, especially outside the band centre.

These results can be applied to both electronic and photonic systems. For
photonic systems we have shown that under realistic assumptions the lasing
threshold of a random laser is close to $\gamma_0$ both in the diffusive and in
the localised regime. Eqs.~(\ref{ergebnis1}) 
and~(\ref{ergebnis2}) thus not only give the distribution of the decay rate of
each individual mode but also a good estimate of the lasing threshold, i.\,e. the
smallest decay rate of a large number of modes.

\begin{acknowledgments}
Valuable discussions with C.\,W.\,J. Beenakker are acknowledged.
\end{acknowledgments}

\bibliographystyle{apsrev}
\bibliography{paper}

\end{document}